\documentclass[aps,pra,reprint,amsfonts,amssymb,superscriptaddress,floatfix,eprint]{revtex4-2}
\usepackage{graphicx}
\usepackage{amsmath}
\usepackage{txfonts}

\DeclareMathOperator\atan{atan}
\DeclareMathOperator\Arg{Arg}
\let\conv=\circledast
\usepackage{dcolumn}
\usepackage{color}
\usepackage{calc}
\usepackage{MnSymbol}
\usepackage{hyperref}
\usepackage{hypernat}
\usepackage{enumitem}
\usepackage{float}
\usepackage{relsize}
\usepackage{varwidth}
\usepackage{tasks}
\usepackage{lipsum}
\usepackage[utf8]{inputenc}
\usepackage[T1]{fontenc}
\hypersetup{
    unicode=true,
    plainpages=false,
    pdftitle={Vortex pair dynamics in three-dimensional homogeneous dipolar superfluids},
    pdfauthor={Srivatsa B. Prasad, Andrew W. Baggaley, and Nick G. Parker},
    pdfsubject={Vortex pair dynamics in three-dimensional homogeneous dipolar superfluids},
    colorlinks=true,
    linkcolor=magenta,
    citecolor=magenta,
    filecolor=black,
    urlcolor=blue
}

\begin{document}

\title{Vortex pair dynamics in three-dimensional homogeneous dipolar superfluids}
\author{Srivatsa B. Prasad}
\email{srivatsa.badariprasad@newcastle.ac.uk}
\affiliation{Joint Quantum Centre Durham–Newcastle, School of Mathematics, Statistics and Physics, Newcastle University,
Newcastle upon Tyne, NE1 7RU, United Kingdom}
\author{Nick G. Parker}
\email{nick.parker@newcastle.ac.uk}
\affiliation{Joint Quantum Centre Durham–Newcastle, School of Mathematics, Statistics and Physics, Newcastle University,
Newcastle upon Tyne, NE1 7RU, United Kingdom}
\author{Andrew W. Baggaley}
\email{andrew.baggaley@newcastle.ac.uk}
\affiliation{Joint Quantum Centre Durham–Newcastle, School of Mathematics, Statistics and Physics, Newcastle University,
Newcastle upon Tyne, NE1 7RU, United Kingdom}

\date{\today}

\begin{abstract}
The static and dynamic properties of vortices in dipolar Bose-Einstein condensates (dBECs) can be considerably modified relative to their nondipolar counterparts by the anisotropic and long-ranged nature of the dipole-dipole interaction. Working in a uniform dBEC, we analyze the structure of single vortices and the dynamics of vortex pairs, investigating the deviations from the nondipolar paradigm. For a straight vortex line, we find that the induced dipolar interaction potential is axially anisotropic when the dipole moments have a nonzero projection orthogonal to the vortex line. This results in a corresponding elongation of the vortex core along this projection as well as an anisotropic superfluid phase and enhanced compressibility in the vicinity of the vortex core. Consequently, the trajectories of like-signed vortex pairs are described by a family of elliptical and oval-like curves rather than the familiar circular orbits. Similarly for opposite-signed vortex pairs their translation speeds along the binormal are found to be dipole interaction-dependent. We expect that these findings will shed light on the underlying mechanisms of many-vortex phenomena in dBECs such as quantum turbulence, vortex reconnections, and vortex lattices.
\end{abstract}

\maketitle
\section{\label{sec:level0}Introduction}
The quantisation of circulation and discretization of vorticity is a striking manifestation of superfluidity in interacting atomic Bose-Einstein condensates (BECs). Whereas classical vortices are characterised by a continuous vorticity field concentrated at the cores of individual vortices, the phase coherence of superfluids results in a vanishing vorticity everywhere but along discrete \textit{topological defect} lines where the superfluid density must necessarily vanish. While superfluid vortices undergo reconnections and annihilations, thereby releasing compressible energy in the process in the form of phonons, the overall vorticity of a non-dissipative superfluid is conserved~\cite{rmp_81_2_647-691_2009}. In the absence of such processes the dynamics of ensembles of quantum vortices in a uniform superfluid background are well-described theoretically by the Biot-Savart law~\cite{prb_31_9_5782-5804_1985, prb_38_8_2398-2417_1988}. Furthermore, if all of the vortices are polarised (anti-)parallel to a given axis and are restricted to move in the plane normal to this axis, they can be modelled as incompressible point vortices~\cite{pnas_27_12_570-575_1941, pnas_27_12_575-577_1941, nuovocimento_6_supp2_279-287_1949}. 
%Quantum vortices have been produced in atomic BECs via a variety of different experimental methods; a nonexhaustive list of examples includes inducing the shedding of vortex pairs of alternating circulation by a moving Gaussian laser beam~\cite{prl_104_16_160401_2010, prl_104_15_150404_2010, prl_117_24_245301_2016}, shaking of the external potential confining the gas~\cite{prl_103_4_045301_2009, nature_539_7627_72-75_2016}, `stirring' of the condensate with an applied laser~\cite{prl_84_5_806-809_2000}, and rotating the external potential itself~\cite{prl_88_1_010405_2001}. 
Ensembles of quantum vortices have been the subject of extensive theoretical and experimental studies. Notably, the ground states of rotating superfluids are generally triangular vortex lattices, accompanied by a spectrum of lattice vibrations in response to external perturbations~\cite{prl_87_6_060403_2001, prl_87_12_120405_2001, prl_91_11_110402_2003, prl_93_19_190401_2004}. Meanwhile, highly agitated superfluids exist in a state of quantum turbulence, a disordered state of vortex lines and phonon excitations of the fluid~\cite{prl_103_4_045301_2009, physrep_622_1-52_2016, nature_539_7627_72-75_2016}. 
%Meanwhile, single vortex lines also undergo helical Kelvin wave excitations. Dynamically, the interactions of vortices and phonons are crucial to the mechanism by which turbulence manifests itself in superfluids and a thorough understanding of vortex dynamics, particularly in the context of single-vortex excitations and vortex-vortex reconnections, is necessary in order to shed light on the finer details of quantum turbulence.
In these macroscopic systems of vortices, the interaction between vortices is key, for example, setting the equilibrium distance between vortices in lattices and driving vortex-vortex reconnections in turbulent systems, which are key to changing the topology of the flow and transfering energy across scales \cite{jltp_128_5-6_167-231_2002, annrevcondmat_11_37-56_2020, avsquantsci_5_2_025601_2023}.  This motivates the importance of understanding the microscopic detail of the vortex-vortex interaction.

These phenomena manifest themselves in strikingly contrasting ways in \textit{dipolar} Bose-Einstein condensates (dBECs) which are composed of lanthanide atoms with large, permanent dipole moments, such as chromium~\cite{prl_94_16_160401_2005}, dysprosium~\cite{prl_107_19_190401_2011, njp_17_045006_2015}, erbium~\cite{prl_108_21_210401_2012}, and europium~\cite{prl_129_2_223401_2022}. Here, assuming that all of the atoms are uniformly polarised along an applied magnetic field, the superfluid properties of the system are strongly modified by the dipole-dipole interaction (DDI) between the atoms. Since the DDI is responsible for \textit{magnetostriction}~\cite{pra_71_3_033618_2005, prl_95_15_150406_2005} -- the energetic preference for dipolar atoms to mutually align themselves along the dipolar axis -- a directional dependence manifests itself in the dynamics of a dBEC including a direction-dependent speed of sound and Landau critical velocity~\cite{prl_106_6_065301_2011, prl_121_3_030401_2018}. Thus, an intricate interplay exists between the DDI and `environmental effects' such as external trapping or rotation. Long before the first experimental detection of vortex structures in dipolar BECs (dBECs) occurred in 2022~\cite{natphys_18_12_1453-1458_2022, crphys_24_S3_1-20_2023}, a considerable body of research has been conducted on single vortex ground states, vortex lattices and the dynamics of vortex ensembles in dBECs subjected to harmonic confinement either in all three dimensions or with a dominant confinement parallel to the axis of vorticity, \textit{viz.} the \textit{quasi-two-dimensional} regime. Notably, the geometry of vortex lattices has been predicted to be strongly dependent on the direction of dipole alignment and strength of the dipole moments, with striped density phases and the possibility of square lattices when the vorticity and the applied magnetic field are not (anti-)parallel~\cite{prl_95_20_200402_2005, pra_75_2_023623_2007, prl_95_20_200403_2005, jphyscondesmatter_29_10_103004_2017, pra_98_2_023610_2018}. Additionally, in these trapped dBECs the cores of individual vortices have been found to be perturbed by the dipolar interaction, leading to elongated cores along the dipolar axis or density rippling associated with the excitation of rotons~\cite{prl_100_24_240403_2008, pra_79_6_063622_2009, jphysconfser_491_1_012025_2014}. Furthermore, the dynamical behaviour of small ensembles of vortices in trapped dBECs has been demonstrated theoretically to differ qualitatively from those of the equivalent nondipolar systems~\cite{prl_111_17_170402_2013, jphysb_47_16_165301_2014, pra_97_4_043614_2018, jlowtempphys_204_1-2_1-11_2021}. We also note that the competing influences of the DDI and externally imposed confinement allows for vortices to be seeded in a trapped dBEC through direct rotation of the magnetic dipole moments by rotating the applied magnetic field~\cite{pra_100_2_023625_2019}, a method which was responsible for the first unambiguous observation of vortices in dBECs in 2021~\cite{natphys_18_12_1453-1458_2022}.

However, relatively little work has been conducted on the physics of vortices in dipolar superfluids in domains that are (roughly) uniform over a significantly sized volume in all three dimensions and which can be effectively realised in optically generated `box traps'~\cite{natphys_17_12_1334-1341_2021, pra_105_6_L061301_2022}. These would serve as experimental platforms to investigate vortex-vortex interactions in a relatively clean system where the interactions of the vortices with boundaries and density inhomogeneities are minimised~\cite{nature_539_7627_72-75_2016}, allowing for the study of quantum turbulence in a regime where the vast majority of theoretical investigations -- including in a dipolar setting~\cite{prl_121_17_174501_2018} -- have been carried out. The possibility of future experiments in this direction motivates the addressing of outstanding questions regarding the elementary properties of vortices in this regime. For instance, it is not presently known whether the anisotropic vortex cores and dipole-mediated dynamics previously predicted in the presence of at least some external confinement are inherent in any dipolar superfluid or if these phenomena are the result of instabilities arising from the interplay between the DDI and the trapping. The work we present here aims to resolve some of these questions and thereby shed light on the properties of single vortex lines and pairs of vortex lines in a uniform $3$D dipolar BEC.

This article is structured as follows. In Section \ref{sec:level1} we provide an overview of the mean-field model employed to study vortices in a three-dimensional, uniform dBEC while, in Sec.~\ref{sec:level2}, we search for stationary states of the system containing a single vortex and examine the ways in which its properties diverge from those of a vortex in a nondipolar BEC. These single-vortex stationary solutions provide insight into the dynamics of vortex pairs, which we first examine for the case where both vortices boast the same circulation in Sec.~\ref{sec:level3} before proceeding to analyse pairs of opposite circulation in Sec.~\ref{sec:level4}. These findings are summarised in Sec.~\ref{sec:level5} along with an outlook to future lines of enquiry in this field.

\section{\label{sec:level1}Formalism}
In this work, we study the structure and dynamics of quantum vortices in dipolar superfluids using the dipolar Gross-Pitaevskii equation (dGPE), which adequately models the behaviour of interacting BECs in the mean-field approximation. Let us consider a single atomic species of mass $m$ and magnetic dipole moment $\mu_{\mathrm{d}}$ which is polarised uniformly by an applied magnetic field parallel to $d$. For this dipolar BEC, the dGPE is given by~\cite{repprogphys_72_12_126401_2009, jphyscondesmatter_29_10_103004_2017, repprogphys_86_02_026401_2023},
\begin{equation}
    i\hbar\partial_t\psi = \left\lbrace-\frac{\hbar^2}{2m}\nabla^2 + g\left[n + 3\varepsilon_{\mathrm{dd}}V_{\mathrm{dd}}(\mathbf{r})\circledast n(\mathbf{r})\right] - \mu\right\rbrace\psi. \label{eq:dgpe}
\end{equation}
Here, the two-body short-ranged interaction strength is $g = 4\pi\hbar^2a_{\mathrm{s}}/m$ with $a_{\mathrm{s}}$ the scattering length of the atom-atom scattering potential, $n = |\psi|^2$ is the atomic density of the condensate, and $\mu$ the chemical potential of the condensate which fixes the normalisation of $\psi$ and satisfies the self-consistency relation,
\begin{equation}
    \mu = \int\mathrm{d}^3r\,\left\lbrace\frac{\hbar^2}{2m}|\nabla\psi|^2 + gn^2 + 3g\varepsilon_{\mathrm{dd}}n(\mathbf{r})\left[V_{\mathrm{dd}}(\mathbf{r})\circledast n(\mathbf{r})\right]\right\rbrace. \label{eq:chempot}
\end{equation}
The parameter $\varepsilon_{\mathrm{dd}} = m\mu_0\mu^2/(12\pi\hbar^2a_{\mathrm{s}})$, where $\mu_0$ is the permeability of free space, serves as an effective dipolar-to-contact interaction ratio that encapsulates the degree to which the dynamics of the superfluid are governed by the dipolar interaction. Furthermore, the two-body dipole-dipole interaction is defined as~\cite{repprogphys_72_12_126401_2009}
\begin{equation}
    V_{\mathrm{dd}}(\mathbf{r}) = \frac{1}{4\pi}\left[\frac{1 - 3(\hat{\mathbf{d}}\cdot\hat{\mathbf{r}})^2}{r^3}\right], \label{eq:ddireal}
\end{equation}
and has a Fourier transform given by
\begin{equation}
    \widetilde{V}_{\mathrm{dd}}(\mathbf{q}) = (\hat{\mathbf{d}}\cdot\hat{\mathbf{q}})^2 - \frac{1}{3}. \label{eq:ddifourier}
\end{equation}
Note that while the global isotropy of the dGPE is broken in the presence of a dipolar interaction, a rotational symmetry is still preserved about the dipolar axis, $\hat{\mathbf{d}}$.

Since no external confining potential enters Eq. (1), its ground state solution is a constant density $n_0$ and the corresponding chemical potential is given by $\mu_{\mathrm{g}} = gn_0\left\lbrace1 + \varepsilon_{\mathrm{dd}}\lim_{q\rightarrow 0}\left[3(\hat{\mathbf{d}}\cdot\hat{\mathbf{q}})^2 - 1\right]\right\rbrace$. Thus, in free space the solutions to Eq. (1) are unstable when $\varepsilon_{\mathrm{dd}} > 1$. While superfluids can still remain stable beyond this threshold, as seen by the experimental realisation of self-stabilised \textit{quantum droplets}~\cite{nature_539_7628_259-262_2016, prx_6_4_041039_2016} and supersolids~\cite{prx_9_1_011051_2019, prx_9_2_021012_2019, prl_122_13_130405_2019} in dipolar BECs in recent years, a theoretical description of this regime requires an extension of the mean-field dGPE to incorporate quantum corrections to the interaction~\cite{ijmpb_20_24_1791-1794_2006, pra_86_6_063609_2012, pra_94_3_033619_2016}. In this work, we explore only the regime $0 < \varepsilon_{\mathrm{dd}} < 1$ for which the mean-field description is sufficient~\cite{repprogphys_86_02_026401_2023}.

%For the remainder of this article, we assume that $V_{\mathrm{T}} = 0$, such that the vacuum solution is indeed given as $\mu_{\mathrm{g}} = gn_0\left[1 + 3\varepsilon_{\mathrm{dd}}\lim_{q\rightarrow 0}\widetilde{V}_{\mathrm{dd}}(\mathbf{q})\right]$. Thus w
We choose to work in \textit{natural units} where 
%all quantities proportional to 
energy is expressed in units of $\mu_{\mathrm{g}}$, length in units of the healing length $\xi=\hbar/\sqrt{m\mu_{\mathrm{g}}}$, time in units of $\hbar/\mu_{\mathrm{g}}$, and $\psi$ in units of $\sqrt{n_0}$~\cite{primer}. Using these natural units, Eq. (1) can be written in dimensionless variables as~\cite{prl_106_6_065301_2011, prl_121_17_174501_2018}
\begin{equation}
    i\partial_t\psi = \left[-\frac{1}{2}\nabla^2 + V_{\mathrm{int}}(\mathbf{r})\conv n(\mathbf{r}) - \mu\right]\psi, \label{eq:dgpedimless}
\end{equation}
where
\begin{equation}
    \widetilde{V}_{\mathrm{int}}(\mathbf{q}) = \frac{1 + 3\varepsilon_{\mathrm{dd}}\widetilde{V}_{\mathrm{dd}}(\mathbf{q})}{1 + 3\varepsilon_{\mathrm{dd}}\lim_{q\rightarrow 0}\widetilde{V}_{\mathrm{dd}}(\mathbf{q})} \label{eq:2bodyfourierdimless}
\end{equation}
is the Fourier transform of the full scaled nonlocal two-body interaction.

It is also helpful to recast the dGPE in a \textit{hydrodynamic form} by writing $\psi$ in the \textit{Madelung form} $\psi = \sqrt{n}e^{iS}$ and interpreting the \textit{superfluid phase} $S$ as a scaled velocity potential for the \textit{superfluid velocity} $\mathbf{v}$ such that $\mathbf{v} = \nabla S$. Under these transformations, the dGPE may be re-expressed as a pair of superfluid hydrodynamic equations~\cite{repprogphys_72_12_126401_2009},
\begin{gather}
    \partial_tn = \nabla\cdot(n\mathbf{v}), \label{eq:continuity} \\
    m\partial_t\mathbf{v} = -\nabla\cdot\left\lbrace\frac{1}{2}m\mathbf{v}^2 + V_{\mathrm{int}}(\mathbf{r})\conv n(\mathbf{r}) - \mu\right\rbrace, \label{eq:euler}
\end{gather}
from which we shall extract considerable intuition about the expected form of numerical solutions to the dGPE. For instance, the quantization of circulation of $\psi$~\cite{rmp_81_2_647-691_2009}, \textit{viz.}
\begin{equation}
    \Gamma = \oint\,\mathrm{d}\mathbf{s}\cdot\mathbf{v} = 2\pi q\,:\,q\in\mathbb{Z}, \label{eq:circulation}
\end{equation}
is inherently a manifestation of the single-valued nature of $\psi$ while the fact that the superfluid density vanishes at the core of a quantum vortex, where $q \neq 0$, is a logical consequence of finding a solution of Eq.~\eqref{eq:continuity} for $n(\mathbf{r})$ where $\mathbf{v}(\mathbf{r})$ is divergent. Hereafter, we consider only vortices with a single quanta of circulation, i.e. $|q| = 1$.

Our results are based on numerical solutions of the dGPE obtained by propagation via the split-step pseudospectral method~\cite{kinetrelatmod_6_1_1-135_2013}, in which the spatial derivatives on the right-hand-side of the dGPE are computed using Fourier-based methods and are propagated separately to the remaining terms. Given that we wish to simulate a system where, far from the vortex cores, the density is approximately constant, the choice of boundary conditions must reflect this. This necessitates the use of distinct spatial grids and corresponding spectral methods for computing the spatial derivatives of $\psi$ depending on the total circulation generated by vortices in the system. These shall be described in greater detail in the relevant sections of this article.

\section{\label{sec:level2}Single Vortex Structures for Orthogonal Dipole Alignments}
In order to best understand vortex pairs, it is important to first examine the properties of individual vortex lines. Surprisingly, vortex lines in uniform $3$D dipolar BECs have received little attention when compared to those in trapped $3$D systems or those that are uniform in \textit{quasi-}$2$\textit{D} limit, a regime where the external confinement along the axis of vorticity is strong enough to render the dynamics of the condensate effectively two-dimensional. We proceed thus to probe the structure of a single vortex in a uniform dipolar BEC and explore the effects of varying the dipolar interaction ratio, $\varepsilon_{\mathrm{dd}}$, and the dipole alignment axis $\hat{\mathbf{d}}$. Without loss of generality we take the vortex to be parallel to the $z$-axis and focus on the two most disparate scenarios, \textit{viz.} dipole alignments along either the $z$- or $x$-axes, which represent alignments parallel or orthogonal to the vorticity, respectively. If the circulation of $\mathbf{v}$ around the boundary of a planar cross-section of a computational cell is nonzero, the boundary conditions in this plane cannot be periodic if the density profile of the condensate is to remain finite since a phase jump between opposite sides of the cell is inevitable. Indeed, any attempt to solve the dGPE in a periodic box with an initial state of nonzero circulation will result in phase defects artificially emerging at the boundaries in order to yield a vortex-neutral state inside the box. Thus, whenever there exists an uneven number of vortices of positive and negative $\Gamma$, we employ \textit{Neumann} (reflecting) boundary conditions in the $x$-$y$ plane~\cite{prl_106_22_224501_2011, prb_88_13_134522_2013}, where the normal derivative of $\psi$ vanishes at the boundaries, and periodic boundary conditions along the $z$-axis. In what follows, the computational grid is of size $(N_x, N_y, N_z) = (513, 513, 2)$ with with equal grid spacings, $\Delta x = \Delta y = \Delta z = 25/256$, along each direction~\footnote{We use only 2 gridpoints along $z$ as uniformity is assumed along this dimension.}. To use pseudospectral methods for computing the spatial derivatives, we also require wavenumber grids of the same size as the position grids but with grid spacings $\Delta k_i = \pi/[(N_i-1)\Delta r_i]$ in the $k_x$-$k_y$ plane and $\Delta k_z = 2\pi/(N_z\Delta z)$. Then, the spatial derivatives may be computed using the discrete cosine transform~\cite{prr_3_1_013283_2021} (DCT)~\footnote{More specifically, to ensure that the same parity conditions are observed at both boundaries along a given axis, we employ the DCT-I transform~\cite{discretecosineandsine}.} while the derivative along the $z$-axis is computed via the discrete Fourier transform (DFT)/fast Fourier transform~\cite{kinetrelatmod_6_1_1-135_2013} (FFT).

\begin{figure*}
\centering
    \includegraphics[width=0.245\textwidth]{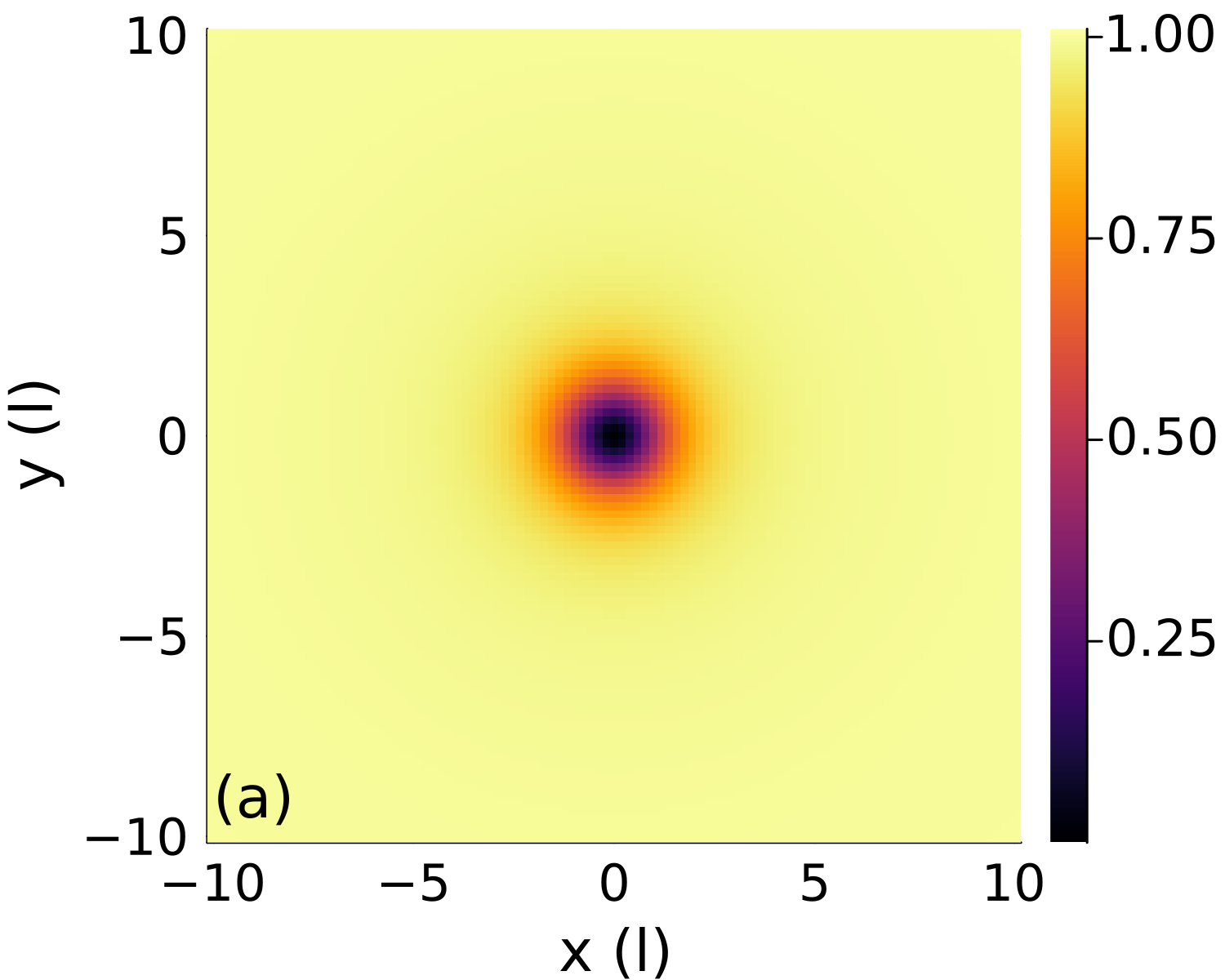}
    \includegraphics[width=0.245\textwidth]{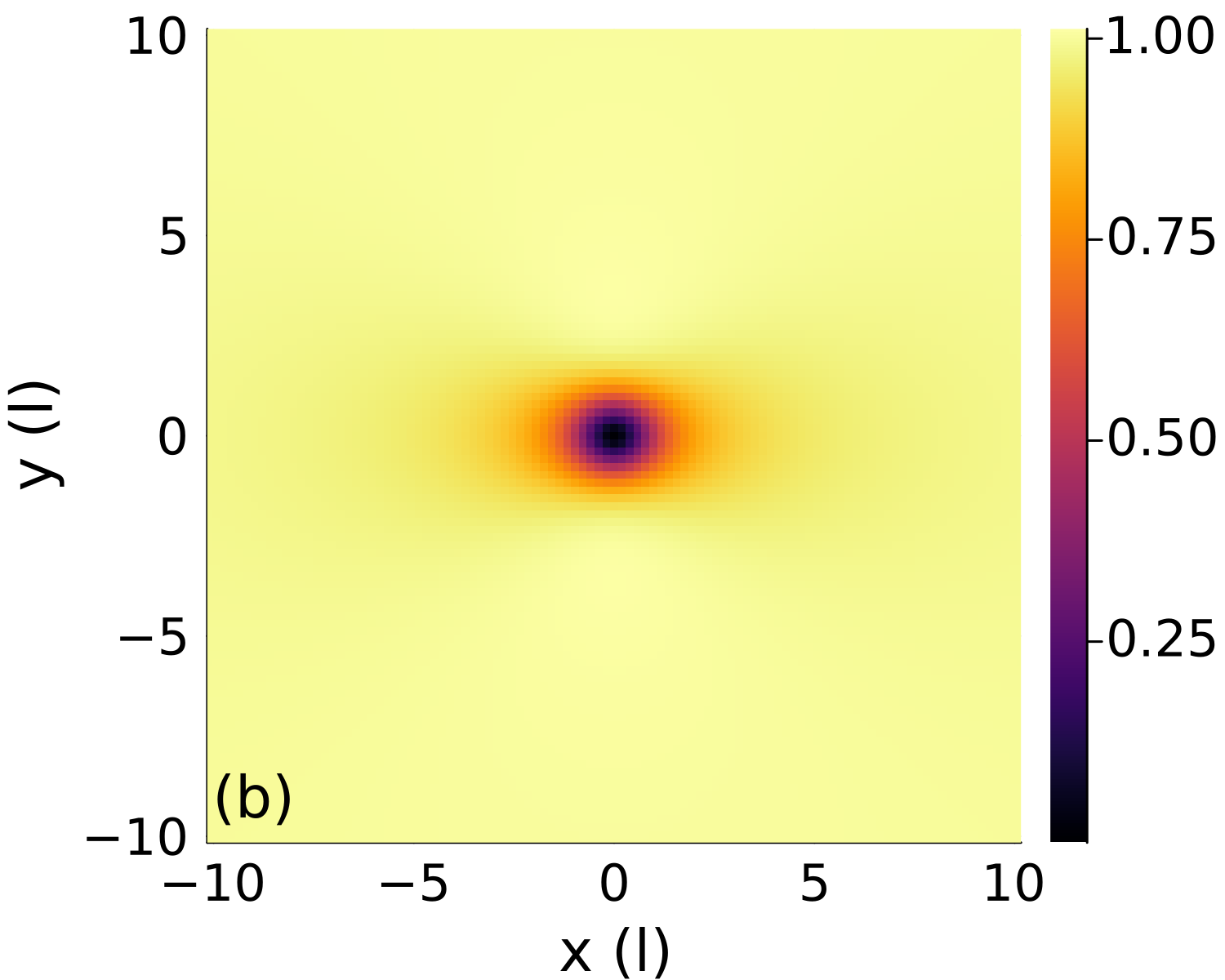}
    \includegraphics[width=0.245\textwidth]{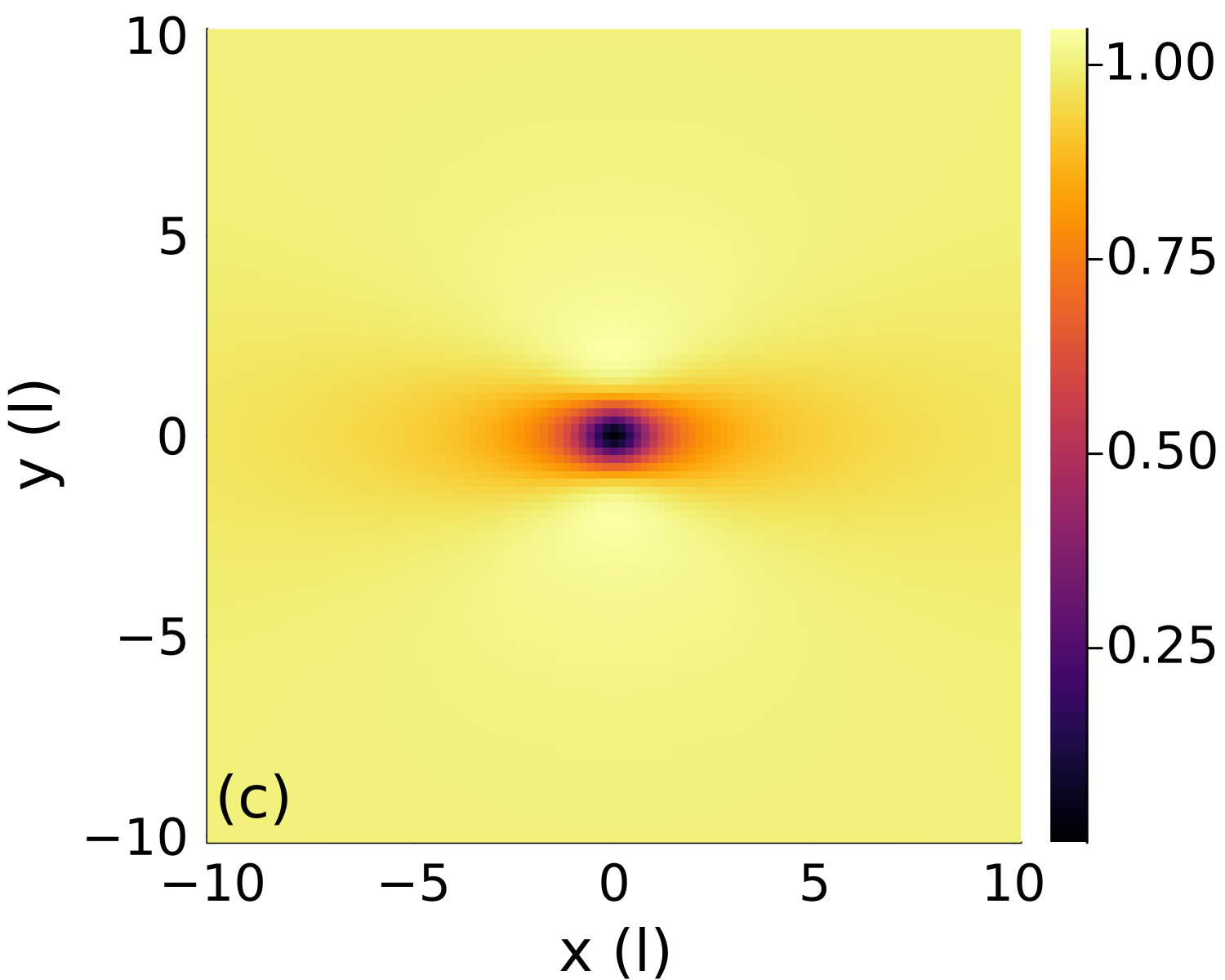}
    \includegraphics[width=0.245\textwidth]{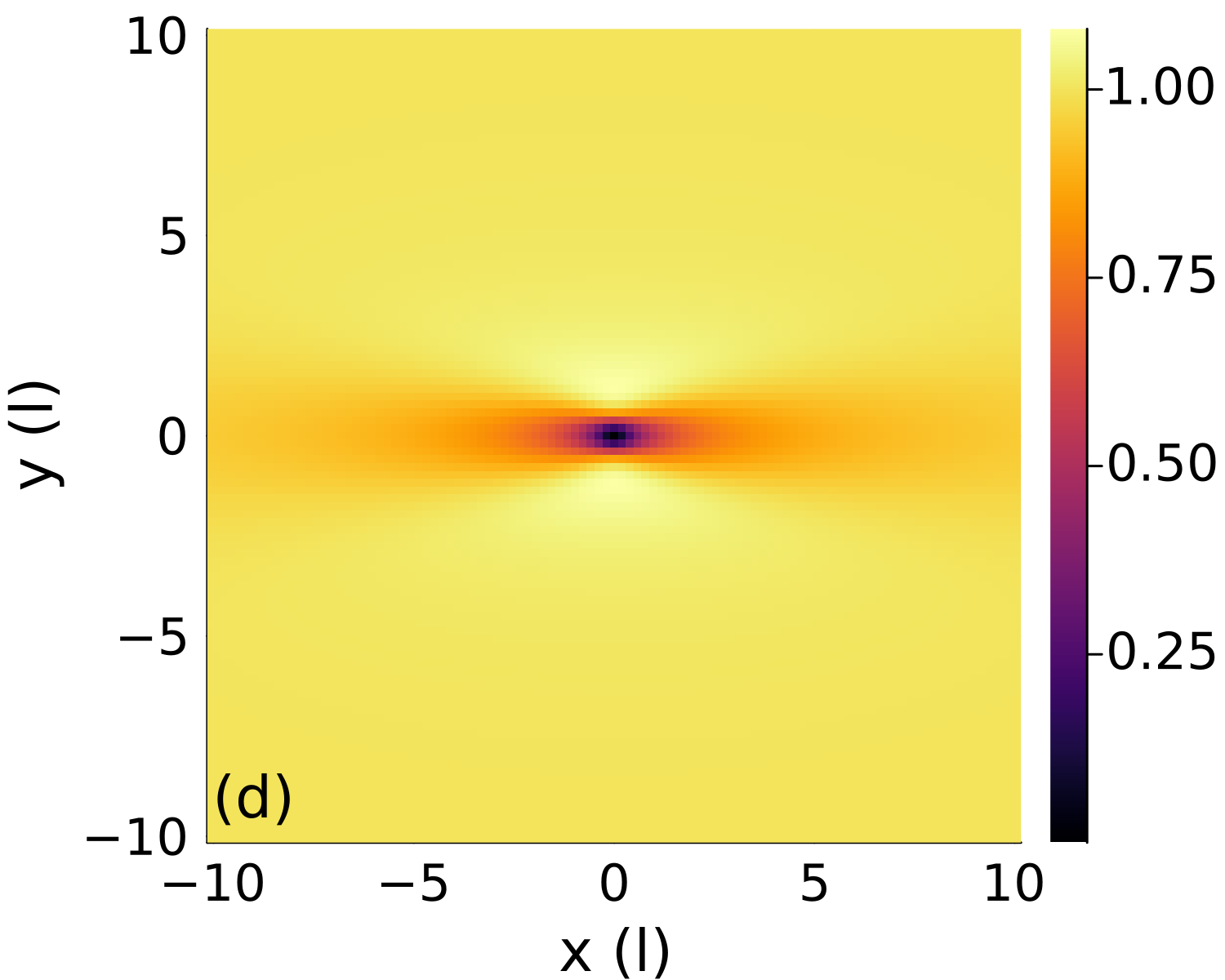}
\caption{Cross-sections of the density profiles, $n(\mathbf{r})$, in the $x$-$y$ plane for a vortex of circulation $2\pi$ where the magnetic field is polarised along the $x$-axis and $\varepsilon_{\mathrm{dd}}$ $=$ $0$ (a), $0.3$ (b), $0.6$ (c), and $0.9$ (d). The density $n(\boldsymbol{\rho}) \sim \rho^2$ as $\rho \rightarrow 0$ and approaches a constant value far from the vortex core. For nonzero $\varepsilon_{\mathrm{dd}}$, the core is elongated along the dipolar axis.}
\label{fig:single_vortex_densities}
\end{figure*}

Let us first consider a dipole alignment parallel to the $z$-axis, for which the condensate solution will be uniform along $z$. Writing $\tilde{n}(\mathbf{k}_{\perp}) = \mathcal{F}\left\lbrace n(\mathbf{r}_{\perp});\,\mathbf{r}_{\perp}\rightarrow \mathbf{k}_{\perp}\right\rbrace$ and evaluating $\left[V_{\mathrm{dd}}\conv n\right](\mathbf{r})$ explicitly yields
\begin{align}
    \left[V_{\mathrm{dd}}\conv n\right](\mathbf{r}) &= \frac{1}{(2\pi)^2}\int\mathrm{d}^3k\,e^{i\mathbf{k}_{\perp}\cdot\mathbf{r}_{\perp}}e^{ik_zz}\tilde{n}(\mathbf{k}_{\perp})\delta(k_z)\left(k_z^2 - \frac{1}{3}\right) \nonumber \\
    &= -n(\mathbf{r})/3, \label{eq:ddiuniform}
\end{align}
implying that the only effect of the dipolar interaction in this regime is to shift the value of the chemical potential depending on the value of $\varepsilon_{\mathrm{dd}}$. Thus, neither the density or phase profile of a straight vortex line are affected by the dipolar interaction if the atomic dipole moments are polarised parallel to the vortex line. In the limit of an infinite domain, the phase of such a vortex at the position $(x, y) = (x_v, y_v)$ is thus given by the nondipolar limit $S(x, y) = q\arctan(y - y_v, x - x_v):\,n\in\mathbb{Z}$. On a grid with reflecting boundaries in the $x$-$y$ plane, the phase is modified at the edges of the grid to account for the boundary condition and it is this modified phase that we initially impose upon the trial state $\psi$ prior to an attempt to obtain a stationary state; a fuller discussion of the nondipolar vortex phase in a reflecting box is provided in Appendix~\ref{sec:level6.1}. We obtain this vortex solution by propagating the dGPE via the normalised gradient method, \textit{viz.} evolving Eq.~\eqref{eq:dgpedimless} with $t \mapsto it$ and renormalizing $\psi$ at each timestep such that the mean value of the density obeys $\langle n\rangle = \langle|\psi|^2\rangle = 1$ as well as reapplying the vortex phase at each timestep. The corresponding density profile is shown in Fig.~\ref{fig:single_vortex_densities} (a). As expected, the density is axially isotropic due to the corresponding isotropy of the nondipolar GPE. Furthermore, matching the properties of a $q = 1$ vortex in a uniform nondipolar BEC, $n(\boldsymbol{\rho})$ is proportional to $\rho^2$ as $\rho$, the distance from the centre of the vortex core, tends to zero, and approaches $\mu/\mu_g$ when $\rho \gg \xi$~\cite{rmp_81_2_647-691_2009}.
%Approaches such as the use of Pad{\'e} approximations have been applied in the literature in an effort to yield more precise characterisations of the density profiles of nondipolar vortices.

By contrast, whenever the dipole polarisation has a nonzero projection orthogonal to the vorticity, we must look to the superfluid hydrodynamic equations to provide some insight about the form of the stationary solutions. Let us consider a dipole moment located at the origin and polarised along the $x$-axis. From the form of the DDI in real space, we observe that its long-ranged interaction of this dipole moment with a second moment is dependent on the angle between the dipole polarisation and the separation axis, $\theta_d$. Since $V_{\mathrm{dd}}(\mathbf{r}) \propto 1 - 3\cos^2\theta_d$, it is energetically preferable for the two dipole moments to minimise the angle $\theta$, an effect referred to as \textit{magnetostriction} and which is responsible for the elongation of trapped dipolar BEC clouds along the polarisation axis. From a superfluid hydrodynamic perspective, a lowered dipolar interaction potential along the axis of the polarisation naturally leads to higher values of the density along this axis in stationary states of the Euler equation, Eq.~\eqref{eq:euler}\cite{pra_61_4_041604r_2000, pra_71_3_033618_2005, prl_95_15_150406_2005}. Via a similar line of reasoning, the anisotropy of the density profile is diminished in the far-field due to the $1/r^3$ dependence of the DDI.

In the particular case of a quantum vortex, its qualitative features can be elucidated through a variation of the preceding arguments. Given that a vortex line is characterised by a localised absence of dipolar superfluid atoms in its core, it may be interpreted as a localised line of dipolar \textit{holes} polarised antiparallel to the atomic dipoles~\cite{prl_100_24_240403_2008} in a manner analogous to the counterparts of electrons in solid-state systems. Given that the generalised form of the DDI for two dipole moments not necessarily parallel to each other is of the form~\cite{repprogphys_72_12_126401_2009, repprogphys_86_02_026401_2023}
\begin{equation}
    V^{(12)}_{\mathrm{dd}}(\mathbf{r}) = \frac{3}{4\pi}\left[\frac{1 - 3(\hat{\mathbf{d}}_1\cdot\hat{r})(\hat{\mathbf{d}}_2\cdot\hat{r})}{r^3}\right], \label{eq:ddiantiparallel}
\end{equation}
it is clear that the energetically preferential alignment for an atomic dipole-vortex hole dipole pair is such that the angle between the applied magnetic field and the mutual separation is $\pi/2$. Thus, the vortex core becomes elongated along the direction of the applied magnetic field as the virtual vortex dipole moments align preferentially along the magnetic field, thereby expelling superfluid atoms in a larger domain along this axis relative to the orthogonal axis. This, we note, is a novel form of the same magnetostrictive effect that induces an elongation of the condensate density envelope of a trapped dipolar BEC along the magnetic axis but is mediated by virtual, rather than real, dipole moments~\cite{pra_73_6_061602r_2006}.

While this core elongation has been discussed extensively in the literature for both three-dimensional trapped~\cite{pra_73_6_061602r_2006, crphys_24_S3_1-20_2023} and quasi-two-dimensional uniform~\cite{prl_111_17_170402_2013, jphysconfser_491_1_012025_2014} dipolar BECs, little attention has been paid to its consequence on the phase profile of the vortex. Focussing on the continuity equation, Eq.~\eqref{eq:continuity} rather than the Euler equation, Eq.~\eqref{eq:euler} we argue that a logical consequence of an anisotropy of the stationary state density profile must be a corresponding anisotropy of the superfluid phase and velocity field. Thus, the normalised gradient descent method is again used to find stationary states of the dGPE with $\hat{\mathbf{d}} = \hat{x}$ where a single vortex is located at the centre of the numerical grid but, while the density is renormalised throughout the propagation, both the (normalised) density and phase are allowed to evolve freely during the search for a stationary state of Eq.~\eqref{eq:dgpedimless}. The results of these simulations are depicted as cross-sections of the density for $\varepsilon_{\mathrm{dd}} = \lbrace 0.3, 0.6, 0.9\rbrace$ in Fig.~\ref{fig:single_vortex_densities} (b), (c), and (d), respectively. Here, it is clear that the vortex core becomes elongated along the dipolar axis for nonzero $\varepsilon_{\mathrm{dd}}$ and that the degree of elongation increases with the dipolar interaction strength. Note that no global density rippling is present in the system, unlike in the quasi-two-dimensional limit where \textit{rotons} can be excited in the vicinity of the vortex core~\cite{prl_111_17_170402_2013, jphysconfser_491_1_012025_2014}; these rotonic ripples are a consequence of external confinement which is absent in our analysis.

\begin{figure*}
    \centering
    \includegraphics[width=\textwidth]{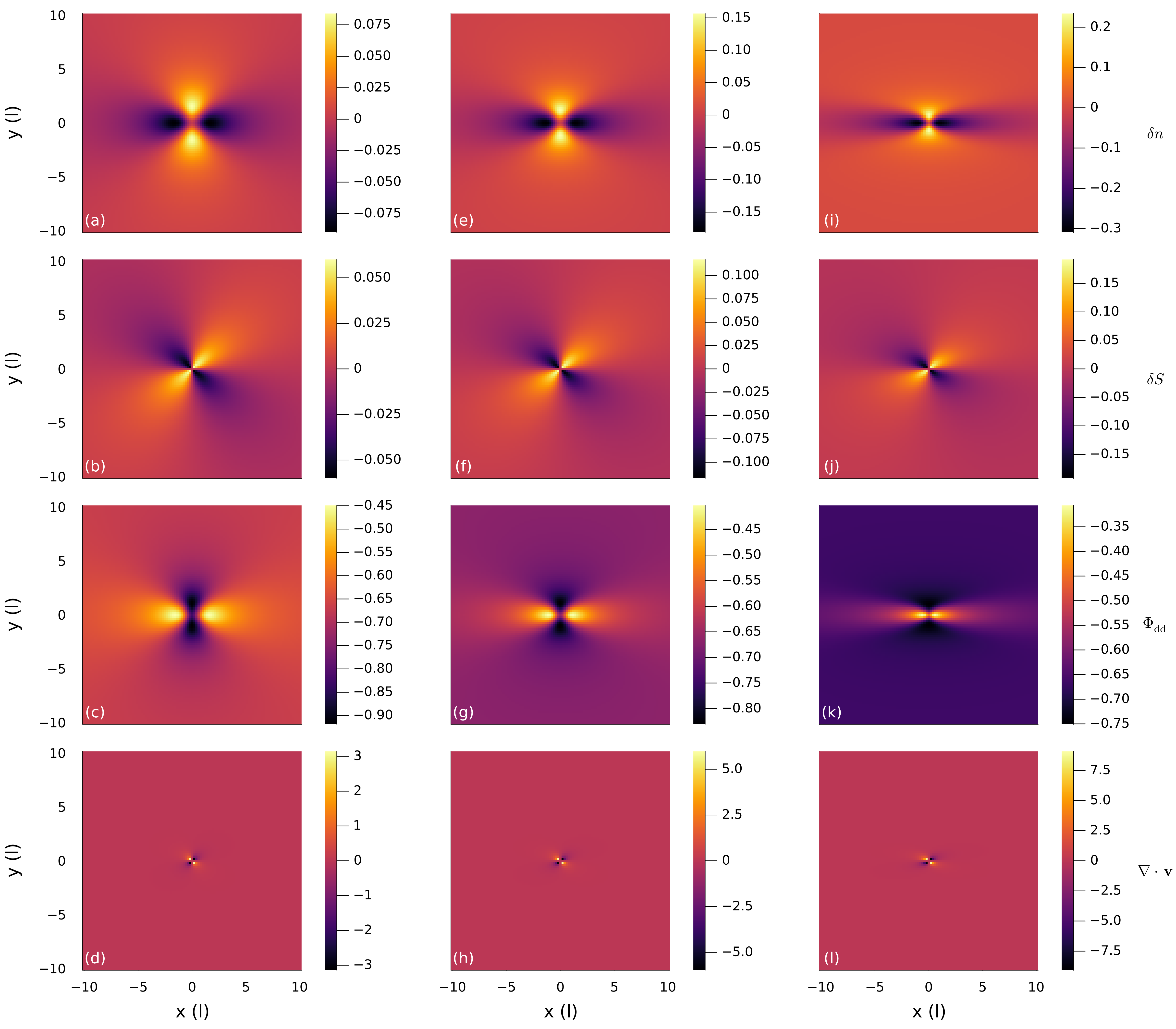}
    \caption{Cross-sections of the axial density anisotropy $\delta n$ (first row), phase anisotropy $\delta S$ (second row), local dipolar interaction energy $\Phi_{\mathrm{dd}}$ (third row), and compressibility $\nabla\cdot\mathbf{v}$ (fourth row) as defined in the main text for $\varepsilon_{\mathrm{dd}}$ $=$ $0.3$ (first column, a--d), $0.6$ (second column, e--h), and $0.9$ (third column, i--l). An inverse correlation between $\delta n$ and $\Phi_{\mathrm{dd}}$ is evident and the maximum magnitude of both $\delta n$ and $\delta S$ increases with $\varepsilon_{\mathrm{dd}}$. N.B. for the sake of clarity, we have set $\nabla\cdot\mathbf{v}$ -- which diverges at the phase singularity -- to zero at $\mathbf{r} = 0$.}
    \label{fig:aniso_twobody_incompress}
\end{figure*}

To gain further insight into these deviations from the nondipolar limit, we also compute
\begin{equation}
    \delta n(\mathbf{r}) = n(\mathbf{r}) - \frac{1}{2\pi}\int_0^{2\pi}\mathrm{d}\varphi\,n(r, \varphi), \label{eq:dendiscrep}
\end{equation}
the discrepancy of the density at a given point from the angle-averaged density at that radius, and $\delta S(\mathbf{r}) = S(\mathbf{r}) - S_0(x + iy)$, the discrepancy of the phase from the incompressible solution, satisfying Neumann boundary conditions, given by Eq.~\eqref{eq:neumannphase} in Appendix~\ref{sec:level6.1}. Furthermore, we evaluate $\nabla\cdot\mathbf{v}$, a measure of the compressibility of the superfluid velocity fields of the stationary solutions, and $\Phi_{\mathrm{dd}}(\mathbf{r}) = \langle V_{\mathrm{dd}}(\mathbf{r})\rangle \equiv n(\mathbf{r}) \circledast V_{\mathrm{dd}}(\mathbf{r})$, the local expectation value of the dipolar interaction potential. Each of these quantities are presented, below, in Fig.~\ref{fig:aniso_twobody_incompress} for the values $\varepsilon_{\mathrm{dd}} = \lbrace 0.3, 0.6, 0.9\rbrace$.

From Fig.~\ref{fig:aniso_twobody_incompress}, it is immediately clear that the density profile anisotropy $\delta n(\mathbf{r})$ in the first row is inversely correlated with $\Phi(\mathbf{r})$ in the third row. In other words, the relative elongation of the vortex cores along the $x$-axis is reflected in the larger values of $\Phi$ along the $x$-axis as the superfluid atoms are ejected from regions in which the local dipolar interaction energy is higher. As we had predicted earlier, the plots of $\delta S$ in the second row demonstrate that the vortex phase is modified by dipolar interaction. Interestingly, both $\delta n$ and $\delta S$ take the form of lobes reminscent of $d$-wave orbitals, a symmetry obeyed by the DDI, with the lobes corresponding to $\delta n$ being out of phase compared to those of $\delta S$ by $\pi/4$. We also observe that the fact that the maximum magnitude of $\delta n$ increases with $\varepsilon_{\text{dd}}$ is accompanied by a similar behaviour exhibited by $\delta S$, which we note is a consequence of $n$ and $S$ mutually satisfying the continuity equation, Eq.~\eqref{eq:continuity}. It is also evident that the phase discrepancy manifests itself in the superfluid velocity field in the form of a nonneglible compressibility in the vicinity of the vortex core in the fourth row of Fig.~\ref{fig:aniso_twobody_incompress}. In particular, this behaviour exhibited by $\delta S$ suggests that the velocity of test particles around the vortex is not uniform for a given radius and that this non-uniformity is enhanced for more strongly dipolar condensates. This, in conjunction with the anisotropy of $\Phi$, bears intriguing consequences for the trajectory of a second vortex with respect to this vortex, which we explore for the remainder of this article.

\section{\label{sec:level3}Same-signed Vortex Pair Dynamics}
In Section~\ref{sec:level2}, it was observed that orthogonal magnetic dipole polarisations had a considerable effect upon the density structure of a vortex and a small, but non-negligible effect upon its phase structure. Just as the necessity of satisfying the superfluid continuity equation in the stationary regime provides an explanation for the interplay between the axial anisotropies of the density and phase, we would expect divergent time-dependent dynamics of small numbers of vortices with orthogonal polarisations. Conversely, as the structure of a straight vortex line was found to be unmodified relative to its nondipolar counterpart when the vorticity is parallel to the dipole polarisation, it would be expected that the dynamics of ensembles of such vortices would likewise be unaffected by a nonzero dipolar interaction strength.

\begin{figure*}
    \centering
    \includegraphics[width=\textwidth]{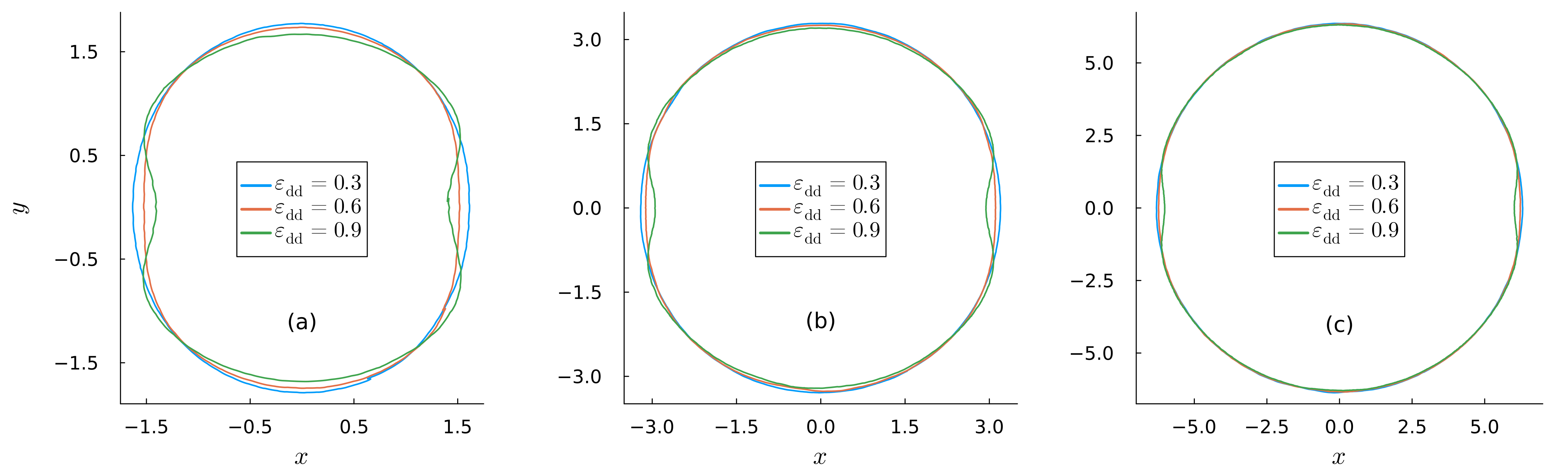}
    \caption{Trajectories of a vortex of circulation $\Gamma = 2\pi$ initially imprinted at the position $x(t=0) = 0$, $y(t=0) = -s/2$, where $s$ $=$ $3.125$ (a), $6.25$ (b), or $12.5$ (c), and accompanied by a vortex of the same circulation located initially at $x(t=0) = 0$, $y(t=0) = s/2$. In each plot, we have $\varepsilon_{\mathrm{dd}}$ $=$ $0.3$ (blue), $0.6$ (red), $0.9$ (black). Due to the redistribution of energy at the start of each simulation the vortices initially increase their mutual separation and we present the trajectories of one closed orbit after a new equilibrium has been attained. Here, an increasing degree of divergence from perfectly circular orbits for larger (smaller) values of $\varepsilon_{\mathrm{dd}}$ ($s$) is apparent.}
    \label{fig:samesignedvortextrajs}
\end{figure*}

Thus, we employ the same numerical grid parameters and mixed Neumann-periodic boundary conditions as in Sec.~\ref{sec:level2} and imprint the phases of two vortices, both of circulation $\Gamma = 2\pi$ at the positions $(x, y) = (0, \pm s/2)$, continuously during a normalised gradient descent propagation of the dGPE until convergence of the stationary solution is obtained. Subsequently, these stationary solutions are propagated in real time with a timestep of $\Delta t = 0.001$; the positions of each vortex are tracked throughout the simulation by identifying points on the grid around which the circulation is approximately $2\pi q,\,q\in\mathbb{Z}$~\cite{pra_81_2_023623_2010, pra_86_1_013635_2012} and then employing a subgrid least-squares interpolation method to further refine the estimated positions of the vortices~\cite{oriordanthesis}. In the nondipolar, incompressible limit, the dynamics of an ensemble of straight, (anti-)parallel vortex lines are governed by the point-vortex model. More specifically, assuming a vorticity about the $z$-axis and indexing each vortex by its position $\mathbf{r}_i$, the positions of each vortex evolve according to the following coupled set of equations,
\begin{equation}
    \dot{\mathbf{r}}_i = \frac{-1}{2\pi}\sum_{j\neq i}\,\Gamma_j\frac{\hat{z}\times(\mathbf{r}_i - \mathbf{r}_j)}{|\mathbf{r}_i - \mathbf{r}_j|^2}. \label{eq:pointvortex}
\end{equation}
For a pair of same-signed vortices of circulation $\Gamma_i = \Gamma_j = 2\pi$ and initial separation $r$, it is easy to show that the solution of Eq.~\eqref{eq:pointvortex} is the vortex pair orbiting their centre of mass at a fixed radius, i.e. in a circular orbit, and moving anticlockwise with an angular velocity $\boldsymbol{\omega} = 2\hat{z}/r^2$. Thus, the solution of the point-vortex model for the two imprinted vortices in our dGPE simulations is that the two vortices are locked in two circular orbits around the centre of mass $\mathbf{r} = 0$ at a radius $s/2$ and period $4\pi s^2$.

However, our numerical simulations demonstrate the existence of orbits whose profiles are distinctly noncircular. In Fig.~\ref{fig:samesignedvortextrajs}, we present plots of the same-signed vortex trajectories with an initial displacement parameter $s = 3.125\xi$ (a), $s = 6.25\xi$ (b), and $s = 12.5\xi$ (c), with $\varepsilon_{\mathrm{dd}} \in \lbrace 0, 0.3, 0.6, 0.9\rbrace$ and $\hat{\mathbf{d}}$ aligned along the $x$-axis throughout the simulations. For the sake of clarity, only one of the two vortex trajectories is plotted for each parameter set; the phase shift of the second vortex with respect to the depicted vortex is always approximately $\pi/2$. We also note that there is an initial transient of phonons released from the vortices at $t=0$ as the solution adjusts, and that this leads to a small initial increase in vortex separation; in Fig.~\ref{fig:samesignedvortextrajs} we plot only the trajectories after the vortices have relaxed into the equilibrium orbits that they occupy thereafter.

In Fig.~\ref{fig:samesignedvortextrajs} the qualitative shape of the trajectory of the vortex at fixed $\varepsilon_{\text{dd}}$ appears to be largely independent of the initial separation as the plots in each panel of the figure are very similar to each other. However, striking differences are evident when focussing on any one of the three panels, thereby fixing the initial vortex-vortex separation, and instead varying $\varepsilon_{\text{dd}}$. When $\varepsilon_{\mathrm{dd}} = 0.3$, the trajectory of the vortex is almost perfectly circular and is characterised by only a small ellipticity with a semi-major axis along $\hat{y}$. A larger degree of ellipticity is seen for $\varepsilon_{\mathrm{dd}} = 0.6$, though, and for $\varepsilon_{\text{dd}} = 0.9$ the vortex trajectories are clearly no longer elliptical but are instead reminiscent of \textit{Cassini ovals}~\cite{moonspencerfieldtheory}.

We interpret these qualitative differences in the trajectories of the same-signed vortices as being due to the anisotropy of the local dipolar interaction energy, $\Phi_{\mathrm{int}}$, experienced by either vortex due to the presence of the other, \textit{cf.} Figs.~\ref{fig:aniso_twobody_incompress} (c), (g), and (k). Whereas the dipolar interaction energy for a dipolar atom located along the $x$-axis relative to a vortex (and thus separated from the vortex line parallel to the dipolar axis) is higher than that for an atom along the $y$-axis, the presence of a second vortex line at a separation parallel to the dipolar axis is energetically favourable compared to the orthogonal case because the second vortex is, as argued in Sec.~\ref{sec:level2}, effectively a line of virtual dipole moments polarised antiparallel to the atomic dipole moments. Thus, the higher kinetic energy arising from the dipole moments being closer together when their separation is parallel to the $x$-axis is counterbalanced by the lower dipole-dipole interaction energy, thereby arising in elliptical and oval-like orbits with a semi-major axis orthogonal to the dipole polarisation. Since the dipole-dipole interaction energy decays with the distance from the vortex core, as evident in the third row of Fig.~\ref{fig:aniso_twobody_incompress}, it is less favourable for orbits to be anisotropic as the mean intervortex distance increases and this is evident in comparing the trajectories in each panel of Fig.~\ref{fig:samesignedvortextrajs} for a given value of $\varepsilon_{\mathrm{dd}}$.

It is important to note that in previous theoretical studies of vortex dynamics in quasi-two-dimensional systems, like-signed vortex pairs were observed to undergo a similar transition in orbit from circular, through elliptical, to oval-like with increasing $\varepsilon_{\mathrm{dd}}$ when the dipole polarisation had a nonzero projection in the $x$-$y$ plane~\cite{prl_111_17_170402_2013}. The same studies in the quasi-$2$D limit found a similar dependence of the dipolar energy cost associated with the existence of a vortex pair on the orientation of the intervortex separation axis, thereby explaining the deviations from the nondipolar point-vortex paradigm. Similarly, a study of same-signed vortex pairs in a harmonically trapped dBEC subject to dissipation predicted different decay rates for the vortex trajectories depending on the initial angle between the vortex separation and the dipolar axis~\cite{jphysb_47_16_165301_2014}. Given our results, we conclude that these discrepancies are an intrinsic feature of same-signed vortex pairs in dBECs and do not arise necessarily from the interplay of external confinement and the DDI.

\section{\label{sec:level4}Vortex Dipole Dynamics}
We now proceed to explore the dynamics of straight, antiparallel vortex lines, i.e. \textit{vortex dipoles}, in an orthogonally polarised dipolar BEC. Returning to the point vortex dynamics of the incompressible limit and setting $\Gamma_1 = -\Gamma_2 = 2\pi$ one finds that $\dot{\mathbf{r}}_{12}$ is now conserved and that $\dot{\mathbf{R}}_{12} = (\hat{z}\times\hat{\mathbf{r}}_{12})/r_{12}$. Thus, the vortex-antivortex pair undergoes translation along the \textit{binormal} axis parallel to $\hat{z}\times\hat{\mathbf{r}}_{12}$ at the constant speed $v = 1/r_{12}$. For instance, for a pair of vortices with circulations $\pm 2\pi$ at the initial positions $(x, y) = [x(0), y(0) \pm s/2]$, the trajectories of the vortices as predicted by the point-vortex model are
\begin{equation}
    [x_i(t), y_i(t)] = \left[x(0) + \frac{t}{s},\,y(0) \pm \frac{s}{2}\right]. \label{eq:vavtraj}
\end{equation}

\begin{figure*}
    \centering
    \includegraphics[width=\textwidth]{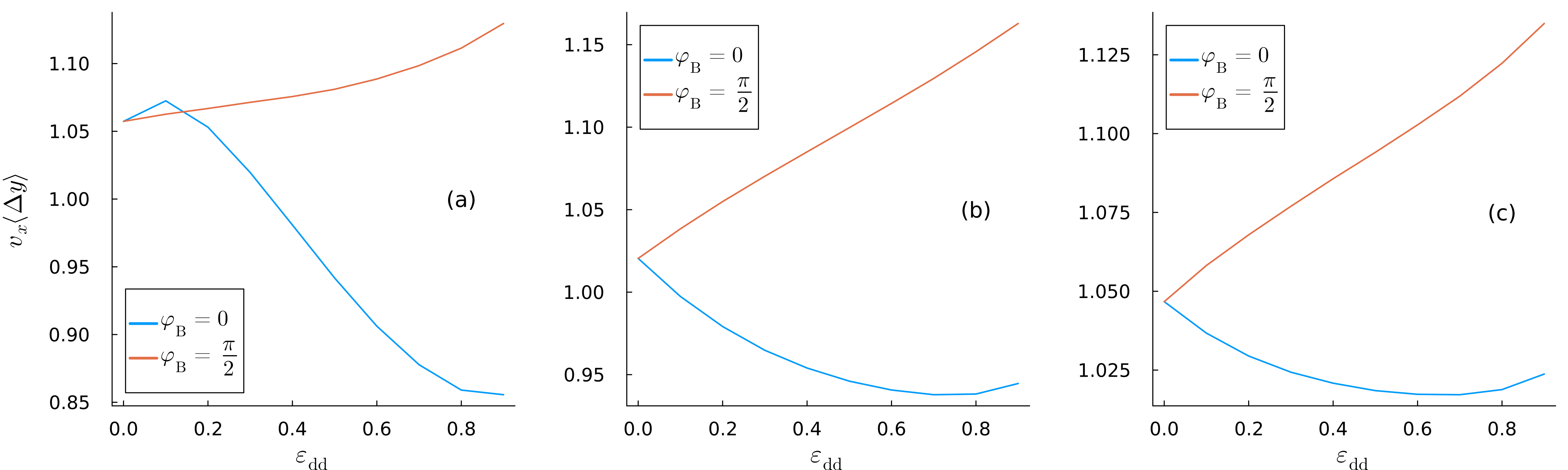}
    \caption{Velocities along the $x$-axis of a vortex-antivortex pair of circulation $\Gamma = \pm 2\pi$ with an initial separation of $d =$ $3.125\xi$ (a), $6.25\xi$ (b), and $12.5\xi$ (c). The vortices initially approach each other slightly before relaxing into a new equilibrium separation due to the effects of the transient energy and thus we ignore the first $25$ units of time. We scale the velocities after this time as $v_x' = v_x(\varepsilon_{\mathrm{dd}})/v_x((\varepsilon_{\mathrm{dd}} = 0)$ and multiply them by the scaled vortex-antivortex separation along the $y$-axis, $\langle\Delta y\rangle' = \langle\Delta y(\varepsilon_{\mathrm{dd}})\rangle/\langle\Delta y(\varepsilon_{\mathrm{dd}} = 0)\rangle$.}
    %{vav_normalised_velocities_and_separations.png}
    \label{fig:vortexdipolevelocities}
\end{figure*}

In the preceding section, however, it was shown that the strongly enhanced \textit{compressibility} of the superfluid flow and, more crucially, the vortex-vortex dipolar interaction cause deviations from the point vortex model that are observable even when the vortex cores are not overlapping. Such deviations would then be detectable in future experiments in box-trapped dipolar BECs~\cite{natphys_17_12_1334-1341_2021, pra_105_6_L061301_2022} where vortex-antivortex pairs are generated by dragging a potential barrier through the system. Hence, we imprint a pair of vortices of circulation $\Gamma = \pm 2\pi$ at the positions $(x, y) = (0, \pm s/2)$ with $s = 6.25\xi$ on a grid with \textit{periodic} boundary conditions in all three directions since the neutral overall circulation of the system allows for a stationary state without the formation of phase defects at the boundaries. Specifically, in this case the dGPE is solved on a $(N_x, N_y, N_z) = (512, 512, 2)$ grid with equal grid spacings in each direction, given by $\Delta x = \Delta y = \Delta z = 25/256$, and the spatial derivatives are computed via FFTs; the wavenumber grids have spacings $\Delta k_i = 2\pi/(N_i\Delta r_i)$. As in the previous section, the true phase of the vortex-antivortex pair is not simply a composition of the phases of the two vortices in the limit of an infinite domain, $\arctan(y, x - s/2) - \arctan(y, x + s/2)$, as we need to account for the effect of the boundaries; the modified phase that we specify in the initial conditions is provided in Appendix~\ref{sec:level6.2}. Similar to our analysis of the same-signed vortex pair, we choose to vary $\varepsilon_{\mathrm{dd}}$ from $0$ to $0.9$ and allow $\hat{\mathbf{d}}$ to be parallel to $\hat{x}$, $\hat{y}$ and $\hat{z}$. The results of these simulations are presented as a plot of $v_x = \dot{x}_i$ as a function of the full range of $\varepsilon_{\mathrm{dd}}$ in Fig.~\ref{fig:vortexdipolevelocities}.

We have found that for $s \gg \xi$, the vortex-antivortex trajectories are still straight lines along the binormal axis except at the earliest times when $s$ adjusts slightly due to the initial transient rearrangement of energy in the simulations. Thus, the plots of the translational velocity in Fig.~\ref{fig:vortexdipolevelocities} are scaled by the separation of the two vortices, $\langle\Delta y\rangle$, \textit{after} an equilibrium has been attained; we note that, in the point-vortex limit, the quantity $v_x\langle\Delta y\rangle$ should be equal to unity. Instead, we find that a strong dipolar dependence exists. For the full range of $\varepsilon_{\mathrm{dd}}$ that we probe, $v_x\langle\Delta y\rangle$ increases monotonically with $\varepsilon_{\mathrm{dd}}$ when the dipole moments are polarised along the $y$-axis, \textit{viz.} parallel to the intervortex separation.

The corresponding behaviour of the vortices when the dipole moments are polarised along the $x$-axis, i.e. translational axis, is less simple. In Fig.~\ref{fig:vortexdipolevelocities} (a), where the vortex separation is initially $3.125$ healing lengths, the translational velocity initially increases with $\varepsilon_{\mathrm{dd}}$ and undergoes a maximum (which exceeds the corresponding value of $v_x\langle\Delta y\rangle$ with the dipole moments polarised along the $y$-axis) before decreasing for larger $\varepsilon_{\mathrm{dd}}$. This is in contrast to the behaviour observed in Figs.~\ref{fig:vortexdipolevelocities} (b) and (c), where the initial separations are $6.25$ and $12.5$ healing lengths, respectively, which we attribute to the effects of a slight overlap of the vortex cores in Fig.~\ref{fig:vortexdipolevelocities} (a). Instead, for the larger separations the scaled translational velocities decrease for larger $\varepsilon_{\mathrm{dd}}$ before attaining a minimum and increasing slightly. The dipolar dependence of the translational velocities are likely due to an interplay between the dipolar correction to the superfluid velocity induced by a single vortex, as elucidated in Sec.~\ref{sec:level2}, and the dipole-dipole interaction between the vortices themselves. We note that our solution scheme approximately preserves the total energy of the system and, due to the point-vortex prediction of constant intervortex separation being upheld by our simulations after the initial mutual approach, the vortex-vortex dipolar energy is also conserved after this initial time. Thus, in comparison to the trajectories of same-signed pairs in Sec.~\ref{sec:level3}, a simple explanation of the vortex-antivortex translation velocity discrepancy in terms of conserving the dipolar interaction energy is less forthcoming. Further investigations of this regime are necessary to clarify these questions. 

It is also pertinent to note that previous theoretical studies of vortex-antivortex pairs in dipolar superfluids have noted a dependence of the critical velocity of vortex-antivortex formation when dragging an obstacle through a dipolar superfluid on the angle between the obstacle velocity and the dipole polarisation, an effect attributed to the anisotropy of the roton dispersion at finite wavenumber resulting from the obstacle's presence~\cite{prl_106_6_065301_2011, eurjphysd_72_3_48_2018}. Other studies in the quasi-two-dimensional regime have suggested that the presence of a transverse dipole-dipole interaction can even suppress the annihilation of a vortex-antivortex pair initially separated closely enough that the equivalent pair in a nondipolar condensate would be annihilated~\cite{prl_111_17_170402_2013}. However, to the best of our knowledge, no prior investigations have uncovered a relation between the vortex-antivortex propagation velocity – as opposed to the critical velocity of an obstacle dragged through a superfluid that is required to nucleate such a pair – and the direction of the dipole polarisation.

\section{\label{sec:level5}Conclusion}
In this work, we have demonstrated that the anisotropy of the magnetic dipolar interaction can fundamentally alter the properties of both the stationary states and dynamics of quantum vortices in a superfluid as compared to the nondipolar paradigm. Whereas prior studies had focussed on scenarios that were then more amenable to experimental probes, such as harmonically trapped systems and systems strongly confined along the axis of vorticity, we have shown that the qualitative deviations that these investigations observed are evident even in a wholly uniform, three-dimensional system. Firstly, the stationary states of a single, straight vortex line are shown to exhibit axial anisotropy in both the density and phase profiles. This is seen to be due to the effective dipolar interaction near the vortex core between the dipolar atoms of the bulk and the virtual oppositely-polarised dipolar holes of the vortex line. While a modification of the phase had been noted in a previous study of single vortices in harmonically trapped dBECs~\cite{pra_73_6_061602r_2006}, this phenomenon had not previously been predicted in the uniform limit. For systems of pairs of vortices, both the like-signed and opposite-signed cases exhibit deviations from the nondipolar regime. In the case of both vortices having the same circulation, we have found that their orbits are described by a family of curves from circles to ovals depending on the strength of the dipole-dipole interaction and their initial separation, a phenomenon we attribute to the effective dipolar interaction between the vortex lines. When the two vortices comprise a vortex dipole, their trajectories remain a translation at constant velocity but these velocities are found to be dipole-dependent; for intervortex separations larger than $\sim 4$ healing lengths, the relationship between the translational velocities, dipole orientation and dipolar interaction strength is found to be essentially universal with vortex-antivortex pairs in a dBEC polarised parallel to the mutual separation moving at a higher velocity than those in a dBEC polarised along the translational axis.

While the direct relevance of our study is for straight vortex lines in a three-dimensional uniform dBEC, it is pertinent to note that the qualitative ramifications of our findings extend beyond this regime to any system with phase defects in a uniform background. For instance, we expect that the nonneglible dependence of the translational velocity of a vortex-antivortex dipole on the direction and strength of the magnetic dipoles is a general feature of soliton-like excitations in dipolar BECs, regardless of the effective dimension of the system, such as Jones-Roberts solitons, vortex rings and rarefaction pulses~\cite{pra_77_4_045601_2008, prl_100_9_090406_2008, eurjphysd_72_3_48_2018}~\footnote{A. Villois and D. Proment, private communication}. Indeed, to the best of our knowledge, the properties of vortex rings have not been studied systematically in a dipolar BEC. Furthermore, given our demonstration that the phase of a vortex is modified by the dipolar interaction when the dipole orientation is not fully parallel to the vortex line, it must be assumed that the superfluid velocity field induced by any two- or three-dimensional phase defect is similarly modified.

These results also raise further questions about the deeper nature of a vortex in a dipolar BEC. For instance, the presence of lobes of compressibility about the vortex core when the dipole moments are orthogonal to the vortex warrants a semi-analytical treatment of the radial and angular dependence of the discrepancy of the phase profile relative to the nondipolar paradigm. We also believe that the strong divergence of the same-signed vortex pair trajectories from the familiar point-vortex predictions warrants an investigation into extending the point-vortex model to account for the dipolar interaction heuristically, thereby allowing for computationally inexpensive predictions of the trajectories of larger ensembles of vortices; we note that such modified point-vortex models have been employed in the studies of BECs subjected to effects such as external trapping, dissipation or the presence of multiple species~\cite{pra_97_2_023617_2018, pra_106_6_063307_2022, pra_107_5_053317_2023}. Additionally, we note that while the effects of transversely polarised magnetic dipoles on the Kelvin wave excitations of single vortex lines have been studied in trapped dBECs~\cite{prl_100_24_240403_2008, njp_11_5_0155612_2009, pra_79_1_013621_2009}, the phase profile modification observed in our study would bear ramifications for the Kelvin wave spectrum of straight vortex lines in uniform dBECs as well. It is also possible that the dipolar dependence of the vortex-antivortex translational velocities bears consequences for reconnection dynamics in three-dimensional dBECs where the translational motion of the vortices induces growth of Kelvin waves; our results would suggest the existence of a directional dependence in quantities such as the reconnection timescale. Finally, while the Tkachenko oscillations of a vortex lattice in a transversely polarised dipolar BEC have been studied theoretically in harmonically trapped systems~\cite{pra_97_4_043614_2018}, our predictions of a large degree of modification to the dynamics of same-signed vortex pairs suggests that the Tkachenko oscillation spectrum would be modified in the uniform background case as well due to the influence of the vortex-vortex dipolar interaction on the lattice dynamics.

\begin{acknowledgments}
This work was funded by Grant No. RPG-2021-108 from the Leverhulme Trust. S. B. P. acknowledges fruitful discussions with Thomas Billam, Brendan Mulkerin and Alberto Villois. This research made use of the Rocket
High Performance Computing service at Newcastle University.
\end{acknowledgments}

\appendix
\section{\label{sec:level6}Vortex Phase Profiles in the Nondipolar Limit}
\subsection{\label{sec:level6.1}Neumann Boundary Conditions in the x-y Plane}
In a $3$D domain with Neumann boundary conditions in the $x$-$y$ plane and periodicity in the $z$-direction, the infinite-domain expression for the phase of a nondipolar quantum vortex at $(x, y) = (x_v, y_v)$, $S(x, y) = n\arctan(y - y_v, x - x_v)$, does not respect the boundary conditions. Previous studies of vortex dynamics in systems with these boundary conditions have avoided any resulting issues by imprinting additional \textit{image vortices} outside the computational domain in such a way that the velocity projections normal to the grid boundaries are roughly zero~\cite{prl_106_22_224501_2011, jfm_808_641-667_2016}.

In this work we adopt an alternative approach. In a sufficiently large but \textit{finite} domain, we assume incompressibility and take the phase to be the harmonic conjugate of the $2$D Poissonian Green function in this domain which, in turn, represents the superfluid equivalent of the stream function. To wit, in a rectangular grid $x \in [0, L_x],\, y \in [0, L_y]$ with reflecting boundaries, the appropriate incompressible phase for a vortex at $(x, y) = (x_v, y_v)$ is given by~\cite{smirnovvol3pt2complexvarssspecfuns, nuclphysb_460_2_397-412_1996}
\begin{equation}
    S_0(w) = \frac{\Gamma}{2\pi}\Arg\left[\frac{\sigma(w - w_v;\omega)\sigma(w + w_v;\omega)}{\sigma(w - \overline{w}_v;\omega)\sigma(w + \overline{w}_v;\omega)}\right], \label{eq:neumannphase}
\end{equation}
where $w = (x + L_x/2) + i(y + L_x/2)$, $w_v = (x_v + L_x/2) + i(y_v + L_x/2)$ $\omega = (Lx, iL_y)$ represents the pair of \textit{half-periods} of the \textit{Weierstrass} $\sigma$ \textit{function},
\begin{gather}
    \sigma[w;(\omega_1,\omega_2)] = \prod_{\substack{m,n=-\infty\\ (m,n)\neq (0,0)}}^{\infty}\left[\left(1 - \frac{w}{\omega_{mn}}\right)\left(\frac{w}{\omega_{mn}} + \frac{w^2}{2\omega_{mn}^2}\right)\right], \nonumber \\
    \omega_{mn} = m\omega_1 + n\omega_2. \label{eq: weierstrasssigma}
\end{gather}

\subsection{\label{sec:level6.2}Periodic Boundary Conditions in the x-y Plane}

By contrast, the imposition of periodic boundary conditions in the $x$-$y$ plane precludes the existence of solutions to the dGPE with a nonzero circulation of $\mathbf{v}$, the superfluid velocity. Thus, we can only consider ensembles of vortex-antivortex pairs and the fundamental building block of their incompressible velocity profiles -- and their corresponding superfluid phases -- is that of a single vortex-antivortex pair for the specified computational grid.

For a single vortex-antivortex dipole where the vortex with circulation $\pm 2\pi$ is located at the position $(x_{\pm}, y_{\pm})$, the effects of the periodic domain boundaries on the corresponding superfluid can be evaluated by considering not only the given pair but an infinite series of image copies comprising a tiling of the $x$-$y$ plane. The solution for the phase is thus the summation of the solution in the infinite-cell limit, $S_{\infty}(\mathbf{r}) = \arctan(y - y_+, x - x_+) - \arctan(y - y_-, x - x_-)$ with $(x_{\pm}, y_{\pm})$ replaced by the coordinates of a given periodic copy of the original vortex pair. For a grid with $x \in [0, L_x)$ and $y \in [0, L_y)$, the solution for the superfluid phase has been provided in the literature by~\cite{prl_112_14_145301_2014},
\begin{align}
    S(x, y) = \sum_{p = -\infty}^{\infty}\biggl\lbrace &\mp\atan\left[\tanh\left(\frac{\pi Y_{\pm}}{L_y} + p\pi\right)\tan\left(\frac{\pi X_{\pm}}{L_y} - \frac{\pi}{2}\right)\right] \nonumber \\
    &+ \pi\left[\Theta(X_+) - \Theta(X_-)\right]\biggr\rbrace - \frac{2\pi(x_+ - x_-)y}{L_xL_y}, \label{eq:weissmcwilliamsphase}
\end{align}
where $X_{\pm} = x - x_{\pm}$ and $Y_{\pm} = y - y_{\pm}$. Taking the gradient of this expression yields an equivalent quantity, up to a constant shift, to the well-established result for the superfluid velocity of a vortex-antivortex dipole in a periodic grid~\cite{physfluidsa_3_5_835-844_1991, pra_101_5_053601_2020}. Following the procedure of Ref.~\cite{prl_112_14_145301_2014}, we replace the infinite series $\sum_{p = -\infty}^{\infty}$ with a finite series $\sum_{p = -P}^P$ for the sake of computational tractability; we have found that no significant suppression of initial `transient' in the real time propagation of the dGPE occurs if $P$ is increased beyond $O(10^1)$. Thus, in our work, we have set $P = 11$.

%\bibliography{bibquantumgases}
\bibliography{main.bbl}
\end{document}